# THE GAME OF TUMBLEWEED IS PSPACE-COMPLETE

LEAR BAHACK

lear.bahack@gmail.com

ABSTRACT. Tumbleweed is a popular two-player perfect-information new territorial game played at the prestigious Mind Sport Olympiad. We define a generalized version of the game, where the board size is arbitrary and so is the possible number of neutral stones.

Our result: the complexity of deciding for a given configuration which of the players has a winning strategy is PSPACE-complete. The proof is by a log-space reduction from a Boolean formula game of T.J. Schaefer, known to be PSPACE-complete.

We embed the non-planar Schaefer game within the planar Tumbleweed board without using proper "bridges", that are impossible due to the board's topology. Instead, our new technique uses a one-move tight race that forces the players to move only according to the protocol of playing the embedded 4-CNF game.

Combinatorial Games, Complexity of Games, Computational Complexity, Game Theory, PSPACE-complete Problems, Two Player Games.

## 1. Introduction

Tumbleweed was invented in 2020 by Mike Zapawa, and was gaining popularity since the early days of the covid-19 pandemic, taking the first prize for the 2020 *Board Game Geek* combinatorial game competition[11]. The logical and visual aesthetics of the game, the simplicity of the rules, the tactic and strategic depth have all contributed to the inclusion of Tumbleweed in the list of abstract games played at the Mind Sport Olympiad[1], alongside the classical games Chess, Hex, Go, Othello and Shogi.

In §2 we define the original and generalized version of Tumbleweed, and the decision problem $WGTC$ which is the subject of this paper.

In §3 we define Schaefer's game[10] and its decision problem $WSPF$. We show that a restriction to uniform 4-CNF formulas keeps the game PSPACE-complete.

In §4 we prove that $WGTC$ is in PSPACE and $WSPF \leq_p WGTC$. The reduction is hard to read, so we progress slowly toward the actual construction, starting from a very high level description.

In §5 we discuss the highlights of our technique, the possibility of a complexity gap between arbitrary configurations and naturally reachable configurations, and suggest a further research question.

## 2. Original and Generalized Tumbleweed

2.1. **Players, Stones, Board and Configurations.** Tumbleweed is played on a hexagonal-tiled board in the shape of a regular hexagon, typically with 6 or 8 cells per side. After a short setup phase, the two players **Red** and **White** play alternately, placing a single stone of their color each move. In this paper we are sympathizing with Red.

Tumbleweed **stones** are red, white or gray (neutral), and have a **strength** parameter ranging from 1 (lowest) to 6 (highest). Cells are either empty or containing a single stone. A **game configuration** consists of the following data: the board size; the player to move next; the color and strength of each non-empty cell.

2.2. **Legal Moves and Winning Condition.** A hexagonal cell (empty or occupied) has 6 directions, corresponding to its 6 edges, from which stones of other cells can be "seen", when all the cells along the line between them are empty. In figure 2.1, cell A sees two red stones, a single white stone, and a single gray stone. Seeing gray stones has no significance. On its turn, Red can either:

(1) Place a red stone at an empty cell seeing at least one red stone.
(2) Replace a white, red or gray stone with a new red stone, as long as the removed stone's strength is (strictly) less than the number of red stones seen from the chosen cell.
(3) Pass (place no new stone).

The strength of a newly placed red stone is defined as the number of red stones it sees. The set of legal moves for White is symmetrically defined.

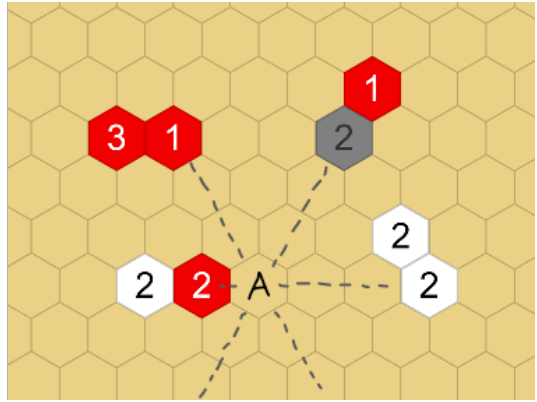

Figure 2.1. Cell A sees two red stones, a single white stone, and a single gray stone. Seeing gray stones has no consequences.

The game ends when both players pass consecutively. Red wins if its **territory** (defined as the number of red stones, regardless of strength) is bigger than White's.

2.3. **Generalizing Tumbleweed.**

***Setup Phase***. Obviously, neither player can make a move in the empty board configuration, rather than pass. Therefore the original tumbleweed game starts with the following setup phase: one player chooses a configuration containing a single 2-strong gray stone at the board's center and a single 1-strong red and white stones, with Red set to move next. Then the other player chooses which color to play (that's the swap rule, intended to keep the opening balanced). The setup phase is irrelevant to this paper.

***Original vs. Generalized Tumbleweed***. In generalized Tumbleweed we consider all possible game configurations, on all possible board sizes, including those that are not reachable from legal setups initial configurations. Essentially it means we allow multiple gray stones.

The rules given here equivalent to the original, but the wordings is different ("stones" instead of "stacks", etc). For the original set of rules and strategic guide see [7].

**Fact 1.** *Starting from any game configuration, each cell can be played on at most 6 times, and the game always ends with a clear winner.*

2.4. **Our Decision Problem.** The Winning Generalized Tumbleweed Configuration problem is:

$$WGTC = \left\{ c \mid \begin{array}{l} c \text{ encodes a generalized Tumbleweed configuration,} \\ \text{with Red to move next and have a forced win.} \end{array} \right\}$$

where a configuration $c$ is of the form:

$$\left\langle k, p, \{s_i\}_{i=1}^{3k^2-3k+1} \right\rangle$$

with $k \in \mathbb{N}_{>0}$ being the size of the board (k cells per edge, and a total of $3k^2 - 3k + 1$ cells on board); $p \in \{Red, White\}$ being the player to move next (in $WGTC$ $p$ is always Red); $s_i \in (\{red, white, gray\} \times \{1, 2, ..., 6\}) \cup \{empty\}$ being the content of the i-th cell. The length of $c$ is then polynomial with respect to the board size.[1]

3. Schaefer's Game

Thomas J. Scheafer has analyzed in his PhD thesis the computational complexity of games based on graphs and Boolean-formulas[10]. Complexity reductions of games and puzzles are often based on Schaefer's fundamental abstract games, either directly or through an intermediate framework such as the Constraint Logic games of Robert A. Hearn's PhD thesis[6].

---

[1]Because there are sparse configurations, this might be false for an alternative encoding, such as listing stones by color, strength and location.

3.1. **Schaefer's Game over Positive CNFs.** Schaefer's $G_{pos}\left(POS\ CNF\right)$ game[10] is played on a positive CNF formula

$$\phi = u_1 \wedge u_2 \wedge \cdots \wedge u_m$$

over an even number of Boolean variables[2] $x_1, \ldots, x_{2n}$. It is sufficient to consider only uniform 4-CNFs formulas, as we show latter. By a *positive* formula we mean the literals are variables without negations, so for suitable indices $1 \leq a_i, b_i, c_i, d_i \leq 2n$ we have

$$\forall 1 \leq i \leq m : u_i = x_{a_i} \vee x_{b_i} \vee x_{c_i} \vee x_{d_i}$$

The players are Red (true) and White (false), with Red going first. On its turn, Red is setting the Boolean value of one of the remaining (unassigned) variables to true, and similarly White to false. After all 2n variables have been assigned, Red wins if $\phi$ is satisfied (iff $\forall 1 \leq i \leq m$, $\{x_{a_i}, x_{b_i}, x_{c_i}, x_{d_i}\}$ contains "true").

3.2. **Complexity of Schaefer's Game.** The computational problem of Winning Schaefer Positive Formulas:

$$WSPF = \{\phi | \text{Red has a winning strategy in } \phi\}$$

is proven to be PSPACE-complete[10]. The reduction built there uses 11-variable clauses (11-CNFs), but to get a tighter reduction we use a recent reasult that shows uniform 4-CNF games are also PSPACE-complete[2]. We may therefore assume that all clauses of a Schaefer game $\phi$ contain exactly 4 distinct variables.

3.3. **Passing Moves.** Using a version of the strategy-stealing argument[3] we prove:

*Claim* 2. Schaefer's game over a positive CNF $\phi$ has the same outcome (in optimal play) when passing moves are allowed. Consequently, the game must end (in optimal play) with a winner, rather than an infinite loop.

*Proof.* By induction. The base case is when $\phi$ has zero clauses and variables, so Red wins immediately either ways. For the general case, we prove that the first player (be it Red only for simplicity) cannot be harmed by making a non-passing move. Assuming the induction hypothesis, this is equivalently means that if Red can win in $\phi$ as a second player, s/he can also win $\phi$ as the first player.

A winning strategy of Red as the second player is modified to a winning strategy as the first player, as follows: first take an arbitrary variable $x_i$. Proceed as if $x_i$ has not been taken, using the strategy of Red as a second player as long as you can. The only move that can't be copied is taking $x_i$ again, and instead Red takes another arbitrary variable. If all variables have been taken, Red is obviously the winner. □

---

[2]In case of an odd number of variables, adding a redundant variable yields an equivalent game.

[3]The strategy-stealing argument[5] is used in many games of symmetric goals, to prove by contradiction that the first player has a winning strategy. According to the argument, the first player can "steal" any winning strategy of the second player, and use it as a first player winning strategy!

In Schaefer's game the goals are not symmetric, but two games are associated with a formula $\phi$, differing by the player moving first. Our version of the argument is of Red as a first player stealing Red's strategy as a second player (similarly for White).

We use claim 2 in §4.5 to be sure altering a certain order of moves is never (strictly) beneficial.

## 4. The computational complexity of Generalized Tumbleweed

First we prove $WGTC$ to be in PSPACE by a generic argument, and then to be PSPACE-hard by proving $WSPF \leq_p WGTC$.

### 4.1. Tumbleweed is in PSPACE.

**Theorem 3.** *$WGTC \in$ PSPACE.*

*Proof.* From fact 1 we know the length of a Tumbleweed game played on a board of size $k$ is at most $6 \cdot (3k^2 - 3k + 1)$ moves long. The game's tree is hence of polynomial depth, which means the winner (when playing optimally) can be computed by the minimax algorithm (essentially a postorder tree traversal). This only requires to keep a polynomial number of configuration at every moment. □

Equivalently, we may use the notion of Alternating Turing Machines and the fact that the alternating polynomial time complexity class APTIME is the same as PSPACE[3].

### 4.2. The Reduction from WSPF to WGTC.
We embed a given positive 4-CNF formula $\phi$ with $m$ clauses and $2n$ variables within a generalized Tumbleweed configuration $c(\phi)$, satisfying $\phi \in WSPF \iff c(\phi) \in WGTC$. The embedding configuration is computable in polynomial time and logarithmic space.

A straightforward technical definition of $c(\phi)$ would be too difficult to comprehend. To better understand the construction, we introduce a virtual tool, external to the configuration: the **protocol** for playing $\phi$ within $c(\phi)$. The protocol is an imaginary restriction of the set of legal Tumbleweed moves, with the following two properties:

(1) Any protocol-violating move is a losing move.
(2) Any sequence of moves (i.e. a path in the game tree) of $\phi$ can be matched to a sequence of protocol moves in $c(\phi)$ and vice versa, with the winner of $\phi$ being also the expected winner of $c(\phi)$. Note that the many sequences in $c(\phi)$ are matched to a single sequence in $\phi$.

Moreover, we break $c(\phi)$ into four types of **gadgets**, that are local arrangements of stones implementing mechanisms that enforce the protocol: **punishments** for the first property, and **incentives** for the second property of the protocol, listed above.

We start with a schematic description of the embedding configuration in §4.3; define the protocol in §4.4; the mechanism of punishments and incentives in §4.5 followed by proofs of the protocol's properties in §4.6. Finally, we construct the gadgets (the actual configuration on board) in §4.7, and formally prove the result of this paper in §4.8.

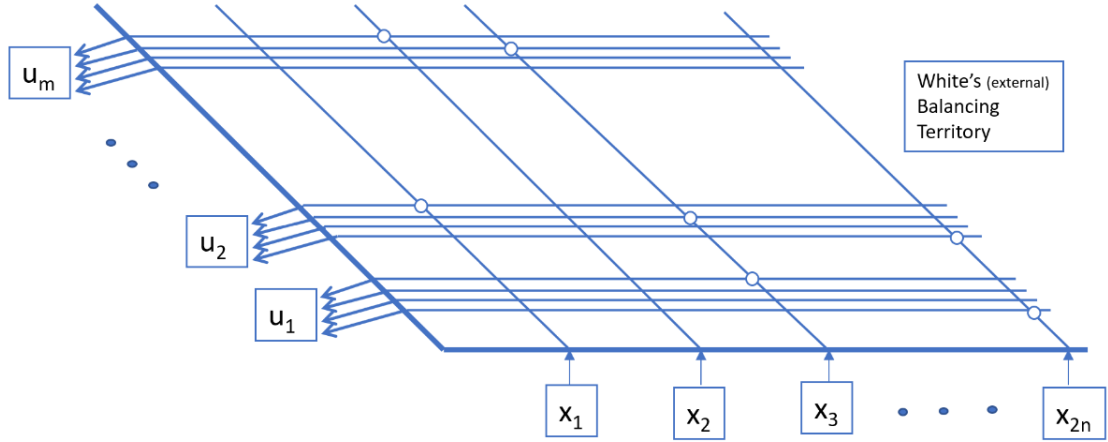

Figure 4.1. Schematic Outline of $c(\phi)$.

4.3. **Schematic Description.** The core of the construction is a grid of horizontal lines and "vertical" lines, that are in fact 120° to the horizon. **Lines gadgets** are one-cell-wide straight strips that are mostly empty, surround by two walls, that are parallel one-cell-wide strips of mostly 6-strong gray stones. Each clause $u_i$ of $\phi$ has 4 parallel horizontal lines, associated with its 4 literals. Each variable $x_j$ of $\phi$ has one associated vertical line. External to the grid there are: a (locked) **white balancing territory**; the **variable gadgets**, located below the bottom edge of the grid, and the **clauses gadgets** located to the left of the grid.

On the left-most edge of the grid, each clause gadget has 4 grid points called **entrances**, connecting the gadget to the 4 horizontal lines of its clause. Each horizontal line of a clause $u_i$ contains a single **incident gadget**, marked by a circle in figure 4.1, at the grid's crossing with the vertical line of the corresponding literal.

4.4. **The Protocol.** There are two phases to the protocol for playing Schaefer's game $\phi$ over the Tumbleweed game $c(\phi)$. Phase II ends with a "boring" game configuration, from which the winner is clearly seen by counting territories: one of Red and White is guaranteed to have an exclusive access to territories with a total size greater than half of the (reachable)[4] board. We take interest only in deciding which player has a winning strategy, as opposed to the highest possible score (or score gap) the winner can guarantee. Therefore, the protocol ends before the Tumbleweed game officially ends.

*Phase I: taking the variables.* Starting with Red, until the last of $x_1, \ldots, x_{2n}$ has been taken (by White, because of parity), the moving player takes one of the remaining variables. $x_i$ is taken by a specific single move in $x_i$'s gadget.

---

[4]Not all cells are reachable: no player can ever get to a 6-strong gray cell, or to a region of empty cells surrounded only by unreachable gray stones and the borders.

*Phase II: evaluating the clauses.* In this phase, Red is securing its exclusive access to the gadgets of all satisfied clauses of $\phi$, i.e. clauses with at least one literal taken by Red. While taking a variable requires a single move only, accessing a clause gadget is a longer procedure, in which the information about the red variable is slowly and securely propagated into the clause. By the end of phase I, each variable $x_i$ taken by Red gives a one red-colored grid point, at the bottom end of the vertical line of $x_i$.

In a sense, Red plays phase II solo: each protocol-legit Red move poses a threat that white has to block by a certain defensive move next. Thus, White is kept busy while Red propagates the influence of its taken variables. When Red is done propagating the information according to the protocol, or is making an off-protocol move, White has no threat to defend from and is free to access any clause gadget that hasn't been accessed yet.

Legit Red moves are of two types:

(1) placing a red stone on the lowest non-red incident on the vertical line of a variable $x_i$ taken by Red.
(2) entering a clause for the first time, through a horizontal line whose incident is already taken by Red.

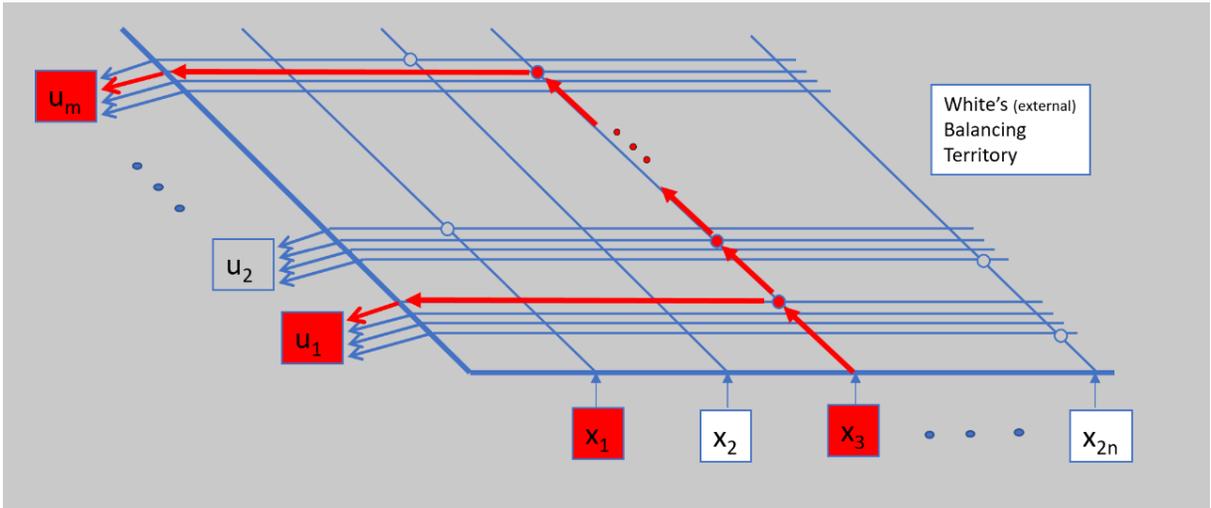

Figure 4.2. a possible configuration during phase 2. Red has yet to access $u_2$, possibly by playing at the incident of $u_2$ with $x_1$.

Phase II ends in one of two scenarios:

(1) Red has accessed all $m$ clauses gadgets, and is therefore the winner.
(2) Red cannot access at least one clause by following protocol moves, and White has therefore accessed at least one clause and is the winner.

4.5. **Territories, Incentives and Punishments.** The grid is a metaphoric skeleton of $c(\phi)$, carrying the metaphoric flesh: the territories. Some territories are meant to make the winner the same as the winner of $\phi$, when the protocol is followed: those are the **incentives**. Other territories are meant to make any protocol violating move a sure losing move: those are the **punishments**.

In this section we set the various territories sizes; the conditions for each incentive and punishment to be activated; prove that any off-protocol move can only harm the player, and that the winner (assuming protocol is followed) is the same winner of Schaefer's game $\phi$. For now, we think of the incentives and punishments as black boxes.

***Incentives.*** A player gets access to a territory of size $\alpha$ for taking each variable $x_i$.

White is (unconditionally) given a **balancing territory** of size $\omega$, already filled with white stones in the initial configuration $c(\phi)$. The balancing territory is needed to make sure White wins if it gets at least one clause, and Red wins if it gets all the clauses.

A clause $u_i$ has a territory of size $\gamma$ given to the player who access $u_i$. Recall that if Red fails to access all clauses in phase II of the protocol, White is going to access at least one of the remaining clauses, and be given its $\gamma$ size territory.

***Punishments.*** A strong violation of phase I is any move that enables the opponent to eventually take more than half of the variables, essentially when the number of remaining variables is odd. A soft violation is any move not taking a variable, when a positive even number of variables remains. Essentially, soft violations are equivalent to passing in Schaefer's game, while insignificantly interchanging the order of some phase I and II moves. Strong violations are punished by a territorial loss of at least $\alpha$ cells to the opponent. Soft violations are not punished, but can only harm the player, according to claim 2.

In Phase II, an off-protocol Red move is punished by allowing White to access a clause territory of size $\gamma$. An off-protocol White move is punished by allowing Red to access one of **White's punishment territories** of size $\beta$, present in all incidents and clauses gadgets.

| Territory type | size | Multiplicity |
|---|---|---|
| | (in cells, per territory) | (number of copies) |
| unreachable cells | irrelevant | |
| $x_i$ gadget ($1 \leq i \leq 2n$) | $\alpha$ | $2n$ |
| white's balancing territory | $\omega$ | $1$ |
| white's punishments (within each clause and incident gadget) | $\beta$ | $5m$ |
| $u_i$ gadget's main territory ($1 \leq i \leq m$) | $\gamma$ | $m$ |
| lines and connectors (i.e. all the remaining reachable cells) | $\lambda$ cells in total | |

TABLE 1. Board cells: types and sizes

***Other cells on board.*** Cells that are not part of incentives or punishing territories, are either unreachable (hence irrelevant to determine the winner) or reachable cells that are part of lines or gadgets entrances. Let $\lambda$ the the total cells number of the latter.

### 4.6. Correctness Claims.

*Claim* 4. If $\alpha > \omega + \lambda + 5m \cdot \beta + m \cdot \gamma$, a player can never gain from off-protocol moves in phase I.

*Proof.* A soft violating move is essentially equivalent to passing in $\phi$, and committing to a certain move in phase II, before phase I is over. According to claim 2, a player can never gain by passing; and obviously committing to a certain move in phase II while is never better than completing phase I first, and then making that same exact move of phase II.

A strong violating move causes the player to lose $\alpha$-size territory to the opponent, which is more than can be gained in the rest of the game, so inevitably any strong violation move is a losing move. □

*Claim* 5. If the protocol was followed in phase I and $\beta > \omega + \lambda + m \cdot \gamma$, any White off-protocol move in phase II is a losing move.

*Proof.* Any White phase II violating move results in a red territory of at least $n \cdot \alpha + \beta$ cells and a white territory of at most $n \cdot \alpha + \omega + \lambda + m \cdot \gamma$ cells. □

*Claim* 6. If $\omega > (m-2) \cdot \gamma + \lambda$, the protocol is followed in phase I, White is following the protocol in phase II and has accessed at least one clause, then White wins.

*Proof.* In such cases, red territory is at most $n \cdot \alpha + (m-1) \cdot \gamma + \lambda$ cells and white territory is at least $n \cdot \alpha + \omega + \gamma$ cells. □

Since any (first) phase II off-protocol move by Red enables White to access at least one clause, we get from claims 5 and 6,

*Claim* 7. If $\omega > (m-2) \cdot \gamma + \lambda$ and the protocol is followed in phase I, any first violation of phase II is a losing move.

At last,

*Claim* 8. If the protocol is followed by both players in both phases and $(m-2) \cdot \gamma + \lambda < \omega < m \cdot \gamma - \lambda$, the winner of the Tumbleweed game is the same as the winner of Schaefer's game (in every matching sequences of moves of $\phi$ and $c(\phi)$).

*Proof.* If White wins in $\phi$, by claim 6 s/he also wins in $c(\phi)$. If Red wins in $\phi$, red territory in $c(\phi)$ will be at least $n \cdot \alpha + m \cdot \gamma$ while white territory at most $n \cdot \alpha + \omega + \lambda$, thus Red wins in $c(\phi)$. □

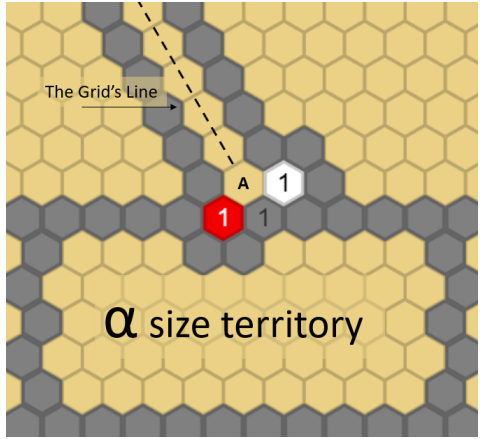

Figure 4.3. A Variable Gadget

4.7. **The Gadgets and Mechanism.** In this section we construct the gadgets that realize the mechanism of incentives and punishments (described in §4.5) on the Tumbleweed board. There are four types of gadgets, and we specify them using diagrams, in which the territories are disproportionately smaller with respect to the constant parts of the gadget. One should keep in mind that the territories in practice are an order of magnitude larger. To get a cleaner view, we use the convention that 6-strong gray stones appear without the number 6, like a solid gray wall.

*Note on the general mechanism.* The construction creates a very tight race: neither of Red and White has a single free move to spare until the end of the protocol. In phase I, lagging in the race causes a loss of $\alpha$-size territory to the opponent (thus losing the game). In phase II, Red (who is the first to move) is keeping White on the toes: every legit Red move is posing a big threat to White, who must spend every move to block the newly posed threat. Once Red makes a single move without a threat, the white beast is released, and is able to access any clause that has not been accessed by Red yet.

Therefore, when examining the possible behaviors of the players in the various gadgets, the reader should keep in mind that neither of the them can make a spare move, and that in phase II the player leading the game (making the first move in the gadgets) is Red.

4.7.1. *Variable*. Both players can take the variable, by making a single move at cell A of figure 4.3, which also serves as a grid point on the vertical line coming out of the variable. When the race is over and the players are free to fill their territories, the player who took the variable will be able to override the 1-strong gray stone, and thus access the territory.

4.7.2. *Incident*. Assume $x_i$ is one of the 4 literals of $u_j$. The incident gadget of the pair $x_i, u_j$ is at the crossing of the vertical line of variable $x_i$ and the horizontal line of $u_i$ associated with $x_i$. The dashed red arrow in figure 4.4 indicates that a red stone might be seen from this direction, exactly when $x_i$ is taken by Red, as well as all the lower incidents of the same vertical line. The solid red arrow means a red stone can always be seen from this direction.

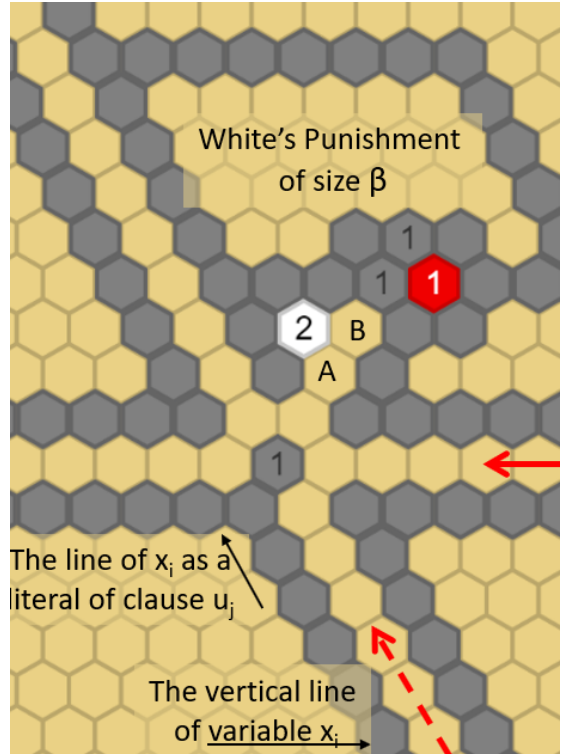

Figure 4.4. An Incident Gadget

Red can legitimately take the incident exactly in the case that a red stone is also seen from below, by overriding the 1-strong gray stone at the crossing. This threatens to let Red access the punishment territory above, unless White respond by playing at either of A,B the following turn. The gadget therefore enables Red to keep White on the toes, and essentially let Red play phase II solo.

4.7.3. *Clause*. There are 4 entrances to a clause gadget, at the crossings of its 4 horizontal lines with the left-most vertical grid line. Two territories are in the gadget: a white's punishment territory of size $\beta$ similar to the ones in incident gadgets, and the clause's territory of size $\gamma$, which will be taken by the first player to access it. As usual, the diagram is disproportional: the territories and the gaps between the 4 horizontal lines are bigger.

A dashed red arrow indicates that a red stone might be seen from this direction. It can only be a stone at the corresponding incident, if the incident is taken by Red. Red can legitimately access the clause if s/he has managed to take at least one of its

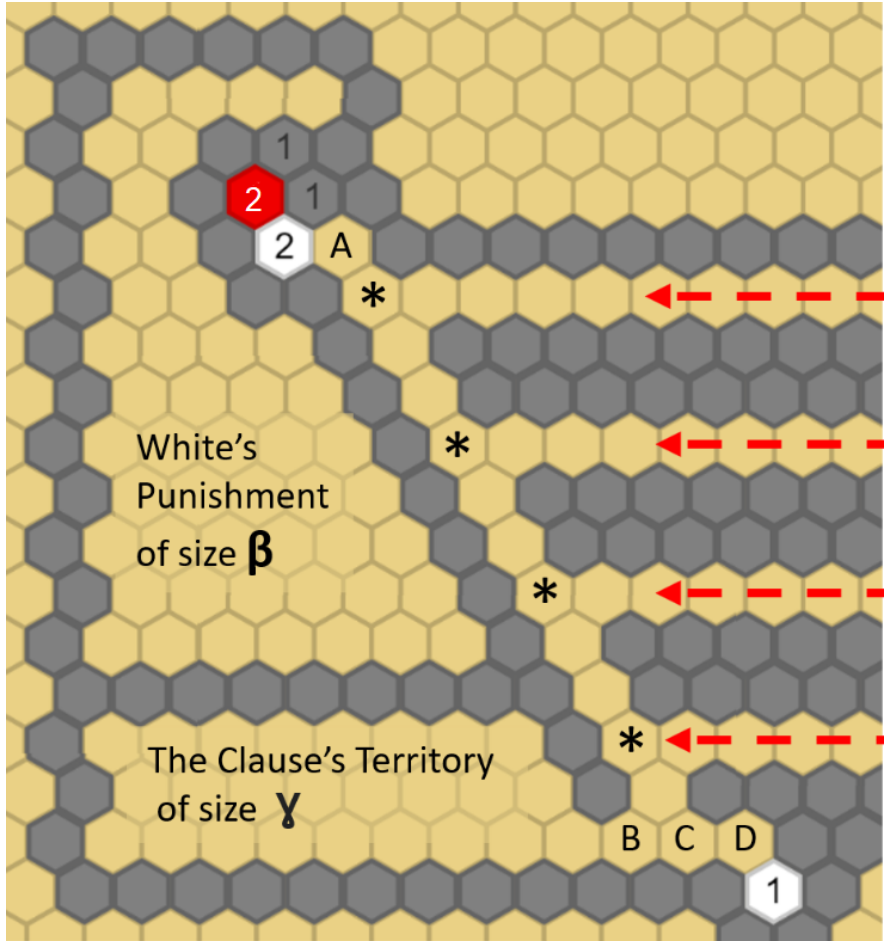

Figure 4.5. A Clause Gadget

4 incidents, by placing a red stone at the entrance (marked with an ∗ in figure 4.5) of the same horizontal line.

Any Red clause-accessing move threatens to access the $\beta$-size punishment territory next, and hence forces White to protect by playing at cell A of figure 4.5. Red's access to the clause's territory is then secured: If White attempts to access the territory, essentially by playing at D, Red responds at C.

In case Red is violating phase II protocol, or has failed to access all clauses after exhausting all possible phase II moves, White has no need to make a threat-blocking move, and is free to play at cell D of a remaining clause. Red won't be able to prevent White from playing next move at B, and thus accessing at least one clause's territory.

4.7.4. ***Vertical and Horizontal Lines***. Lines are the fourth and simplest type of gadgets. We have already seen the bottom ends of vertical lines (connecting to the variables) gadget), the left-most ends of horizontal lines (entering the clauses), and the incident crossings of vertical and horizontal lines.

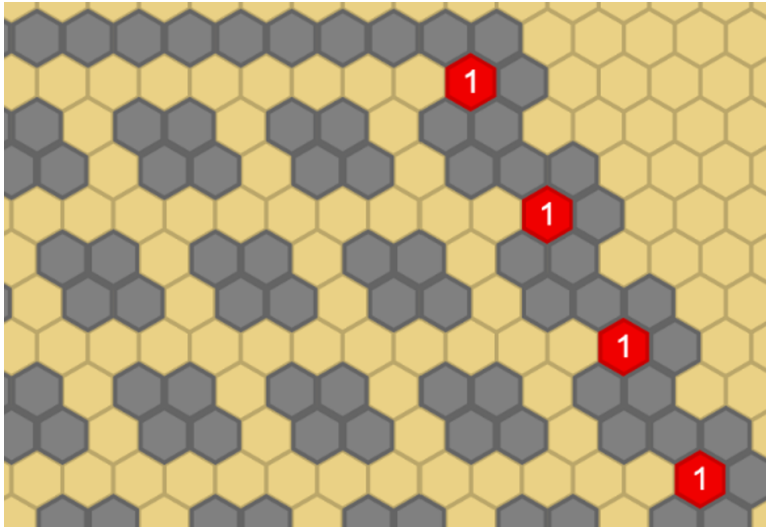

Figure 4.6. Top and Right ends of Vertical and Horizontal lines

The non-incident crossings are simply empty crossings, and the vertical lines end up along the wall of the top horizontal line, as one would expect. The horizontal lines, however, end with a deadlocked red stone past the rightmost edge of the grid. The red stones are needed for the incident gadgets mechanism to work.

Figure 4.6 shows the upper right corner of the grid, assuming there are no incidents there. Note that the gaps between adjacent vertical and horizontal lines are disproportionately small in the diagram.

### 4.8. **Generalized Tumbleweed is PSPACE-Complete.**

*Claim* 9. Given a positive 4-CNF $\phi$ with $m$ clauses and $2n$ variables, a generalized Tumbleweed configuration $c(\phi)$ on a board with $O(m^8)$ cells that is equivalent to Schaefer's game over $\phi$, can be computed in a polynomial time.

*Proof.* We may assume $2n = 4m$, or else we can add or omit variables that do not appear in $\phi$, while keeping the formulas equivalent. We set:

$$\begin{aligned}\lambda &= 256 \cdot m^5 \\ \gamma &= 2\lambda \\ \omega &= 2(m-2)\lambda \\ \beta &= 4m\lambda \\ \alpha &= 24m^2\lambda\end{aligned}$$

and observe that the inequality conditions of all the correctness claims of §4.6 hold true. We set the gap between consecutive parallel (vertical and horizontal) grid lines to be $32m^3$, to get enough room for the White's punishment territories of size $\beta = (32m^3)^2$ each. To make the calculation simple, we ignore the $o(\lambda)$ cells that form the boundaries of the incident gadget territories ($\beta$ still satisfies the condition of claim 5).

The territories of sizes $\alpha, \gamma, \omega$ are all external to the grid, and their total number of cells is less than $(m+1)\alpha = O(m^8)$. Embedding the gadgets within the smallest possible hexagonal Tumbleweed board is obviously possible in polynomial time. $\square$

Finally, from the above claim and theorem 3, we get the result of this paper:

**Theorem 10.** *The decision problem WGCT is PSPACE-complete.*

*Remark.* WGCT is in fact log-space complete in PSPACE: Schaefer's game is log-space complete in PSPACE[10], and our reduction can be computed in logarithmic space as well.

## 5. Conclusion

5.1. **Virtual Bridges using Races: the Highlight of our Technique.** While abstract two-player perfect-information games are usually played on a two-dimensional board, Schaefer's game on positive CNFs as well as similar logic or graph based PSPACE-complete games is non-planar. A central challenge of construction a reduction to a board game, is to propagate information along crossing lines without interference. Gadgets designed to support information crossing are called **bridges** in the literature, and are known to be difficult to construct. Robert A. Hearn and Erik D. Demaine have built a planar infrastructure for game complexity reductions[6], with the goal of eliminating the need of bridges. Regrettably, it seems their game cannot be directly simulated in Tumbleweed.

A **proper bridge**, in which interference to the desired information propagation is unconditionally impossible, cannot exist in Tumbleweed: The hexagonal board has no cross-cuts, hence a player capable of crossing a line is inevitable capable to "contaminate" the line as well (note that in Tumbleweed a player can never be harmed by an extra cell of its color). We have overcome this major obstacle using **races**: the gadgets and grid crossings are designed to be **virtual bridges**, in which interfering with the desired flow of information is possible but causes the player to lag by at least one critical move, which is enough to secure the virtual bridges.

Virtual Bridges using races is the highlight of our technique, and we believe the same technique can be applied to other games in which true bridges are hard or impossible to construct.

5.2. **The Implication of our Result.** Complexity reductions of games usually contain very different configurations than those occurring naturally in normal play. One might wonder if there is a gap between the complexity of solving arbitrary game positions, and solving reachable game positions. Theoretically, it might be the case that the set of configuration reachable in normal / optimal play is so restricted that its complexity class is smaller.

The games of Checkers[9], Chess[4] and Go[8] are likely examples of this complexity gap: While in practice the length of the games is linear with respect to the board size (which implies solvability in PSPACE, see theorem 3), there are absurd configurations in which the game length is exponential, that can be used to show the games are EXPTIME-hard. In Go for instance, an exponentially long game inevitably requires an exponential number of Ko moves. Any Go player knows such a complex system of Koes is extremely unlikely to be reachable in normal or optimal play.

In the case of Tumbleweed, the author believes there is no complexity gap between natural and arbitrary game configurations.

5.3. **A Reduction Without Gray Stones?** An open question is whether a reduction from Schaefer's game to Tumbleweed is possible without using gray stones. We believe the answer is yes, using the same technique of virtual bridges by races.

The gray stones enable us to efficiently set the sizes of incentives and punishments. Without them, upper and lower bounding the scores in case of protocol violation is prone to be challenging, yet possible.

The author would like to thank Alek Erickson and Michael Amundsen for introducing him to the wonderful Tumbleweed game.